\documentclass[12pt,preprcluint]{aastex}
\shorttitle{Effects of Inhomogeneities withing Colliding Flows}
\shortauthors{Carroll-Nellenback, Frank, Heitsch}
\providecommand{\e}[1]{\ensuremath{\times 10^{#1}}}
\begin{document}
\title{The Effects of Inhomogeneities within Colliding Flows on the Formation and Evolution of Molecular Clouds}

\author{Jonathan J. Carroll-Nellenback\altaffilmark{1}, Adam Frank\altaffilmark{1}, Fabian Heitsch\altaffilmark{2}}
\altaffiltext{1}{Department of Physics and Astronomy, University of Rochester, Rochester, NY 14620}
\altaffiltext{2}{Department of Physics and Astronomy, University of North Carolina Chapel Hill, Chapel Hill, NC 27599}
\email{johannjc@pas.rochester.edu}

\begin{abstract}
Observational evidence from local star-forming regions mandates that star formation occurs shortly after, or
even during, molecular cloud formation. Models of the formation of molecular clouds in large-scale converging
flows have identified the physical mechanisms driving the necessary rapid fragmentation. They also point to global 
gravitational collapse driving supersonic turbulence in molecular clouds. Previous cloud formation models
have focused on turbulence generation, gravitational collapse, magnetic fields, and feedback. Here, we explore
the effect of structure in the flow on the resulting clouds and the ensuing gravitational collapse. We compare two
extreme cases, one with a collision between two smooth streams, and one with streams containing small clumps.
We find that structured converging flows lead to a delay of local gravitational collapse (``star formation").
Thus, more gas has time to accumulate, eventually leading to a strong global collapse, and thus to a 
high star formation rate. Uniform converging flows fragment hydrodynamically early on, leading to the rapid 
onset of local gravitational collapse and an overall low sink formation rate.
 \end{abstract}


\section{Introduction}

The concept of flow-driven cloud formation \citep{Vazquez-Semadeni1995,Ballesteros-Paredes1999,Hartmann2001} can explain
two observational constraints on how molecular clouds form stars, derived from local star-forming regions: first, 
all local molecular clouds are observed to be forming stars, and second, the stellar age spreads are on the order of $1-2$~Myr, 
several times shorter than cloud crossing times 
\citep[see summary in ] {Hartmann2001,Ballesteros-ParedesHartmann2007}. 
The first constraint suggests that star formation sets in immediately (or even during) molecular cloud formation, 
and that the second constraint is trivially fulfilled in a scenario where the clouds themselves form in large-scale 
``converging" flows. The immediate (``rapid") onset of star formation in the forming clouds and the fact that the 
star formation efficiency is only a few percent \citep{Evansetal2009} mandates that the clouds are highly structured: 
local collapse must set in before global collapse can overwhelm the dynamics. 

The notion of cloud formation in converging flows has led to a series of numerical experiments investigating the 
physical processes relevant for the rapid fragmentation and for the control of the star formation efficiency. 
There is agreement across the models (despite different initial and boundary conditions) on the following results: 
(1) Rapid fragmentation is induced by strong radiative losses during the flow collision (possibly even by thermal 
instability if the clouds form from atomic gas), and by dynamical instabilities 
\citep{Hueckstaedt2003,Audit2005,Vazquez-Semadeni2006,Heitsch2008b}
(2) Turbulence in molecular clouds is a natural result of the dynamical instabilities during 
the cloud formation, and is driven by global gravitational collapse at later stages of the cloud  evolution 
\citep{Vazquez-Semadeni2007}. (3) Strong, non-linear density contrasts can also be driven by 
self-gravity in finite clouds, due to geometry (or ``edge") effects \citep{BurkertHartmann2004}. 
(4) Although the rapid fragmentation can keep the star formation efficiency low, eventually, feedback or cloud 
dispersal is needed to prevent a large percentage of the gas to participate in gravitational collapse 
\citep{Vazquez-Semadeni2010}.

The most obvious difference in the simulation results is the morphology of the forming clouds. All models use some 
mechanism to break the symmetry in the colliding flows -- otherwise, the flow collision would result in a 
plane-parallel shock. Models with small-scale perturbations (``noise") in the velocities tend to lead to extremely 
flattened clouds with a strong ring due to the gravitational edge effect 
\citep{BurkertHartmann2004,HartmannBurkert2007} in finite, sheet-like clouds. If 
the velocity perturbations are imposed on larger scales (e.g as a turbulent power spectrum), or 
if the collision interface between the two flows is perturbed, turbulent structures form that are not 
necessarily coherent when viewed from different directions \citep{Heitsch2009}. 

To understand better the effect of initial conditions on the clouds forming in the flow collisions, we present a simple experiment, 
comparing two (otherwise identical) cloud formation simulations, one with a smooth inflow, and one with a clumpy 
inflow of identical mass. The experiment is also motivated by the analysis of \citet{Pringle2001}, suggesting 
that cloud formation out of warm atomic gas would require time and length scales too large to be feasible (see also \citet{McKeeOstriker}). This problem is solved by the realization that the flow collision is three-dimensional, 
allowing gravitational collapse and accretion along the dimensions perpendicular to the flows, and thus circumventing 
the one-dimensional limit on column density accumulation \citep{Heitsch2008a}. Yet, \citet{Pringle2001} suggest that clumpy  flows 
could reduce the accumulation and molecule formation time scale (especially if the clumps are already molecular). 
Here, we will test, what effects a clumpy flow has on the resulting cloud and star formation process.

\section{Method, Initial Conditions, and Parameters} \label{numerical_model}

To model a finite molecular cloud forming in a collision of two flows, we use the adaptive-mesh-refinement 
code AstroBEAR 2.0 to solve the equations of hydrodynamics including self-gravity and equilibrium cooling. 
For a detailed discussion of AstroBEAR, see \citet{AstroBear2}. Poisson's equation is 
solved with HYPRE \citep{hypre}. We used a non-split CTU integrator following \citet{GardinerStone2008}, and the 
sink particle implementation discussed by \citet{federrath2010b}.

We performed two simulations of $40$~pc diameter flows with a mean density $n=1.0$~cm$^{-3}$, colliding 
head-on at $v_0=8.25$ km~s$^{-1}$ for a period of 30~Myr.  The flows collide in the y-z plane within 
a box that is $62.5\times100\times100$~pc$^3$ in size.  We use a base grid of $40\times64\times64$ cells 
with $5$ additional levels of refinement for an effective resolution of $1280\times2048\times2048$ and a physical cell 
size of $\approx 0.05$~pc.  The flows combine to give a mass flux of $\dot{M}=665$~M$_\sun$~Myr$^{-1}$ and 
a ram pressure of $P_{\mbox{ram}}=10472$ K~cm$^{-3}$. We used an ideal equation of state at $\gamma=5/3$, with 
a mean particle mass $\chi=1.27$ and a parametrized cooling function $\mathcal{S}$ that includes heating 
terms consistent with \cite{InoueInutsuka08} though modified to give lower temperatures $10$~K at higher 
densities ($n > 10^3$~cm$^{-3}$) to account for UV shielding (Ryan \& Heitsch in prep).  
\begin{equation}
\begin{array}{l}
\mathcal{S} = n(-\Gamma+n\Lambda) \mbox{erg cm}^{-3}\mbox{s}^{-1} \\
\Gamma=2\e{-26} \\
\frac{\Lambda}{\Gamma}=1.0\e{7} \exp{ \left ( \frac{-118400}{T+1000} \right ) }+1.4\e{-2} \sqrt{T} 
\exp{ \left ( \frac{-22.75}{\max{ \left [ 1.0,T-4 \right ] } } \right) }
\end{array}
\end{equation}

\begin{figure}
\includegraphics[width=\textwidth]{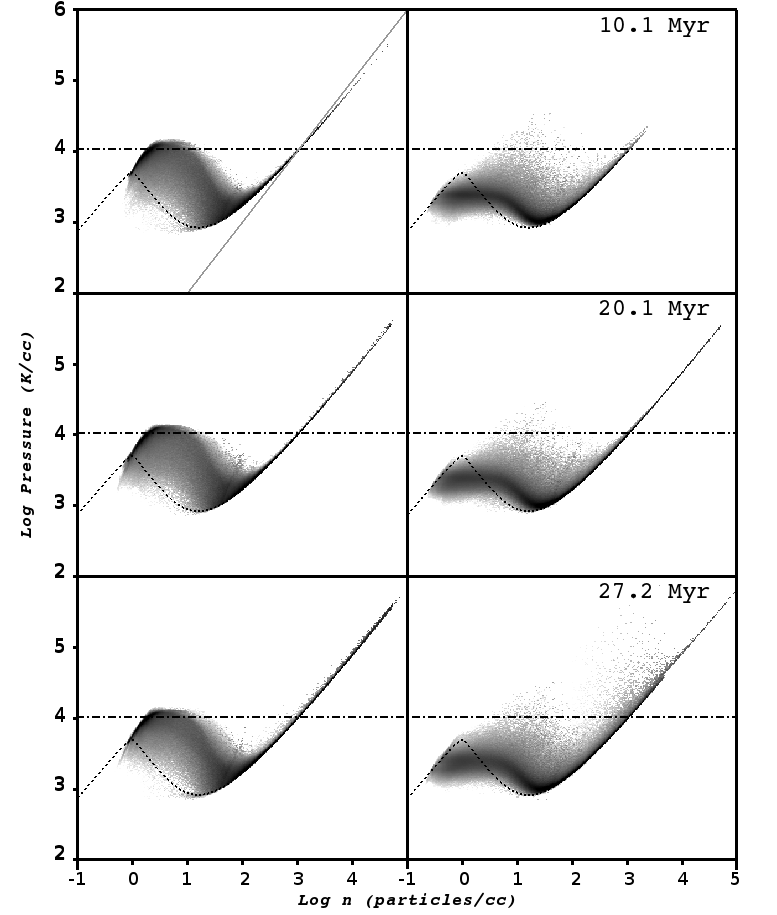}
\caption{Density-weighted joint probability distribution function for density vs. pressure for the 
Smooth run (left) and Clumpy run (right), at $10.1,20.1$, and $27.2$~Myr.  The dot-dashed line 
indicates the ram pressure of the flows, and the diagonal line in the upper left plot shows $T\equiv10$~K.}
\label{pdfs}
\end{figure}

\paragraph{}
The combined heating and cooling results in a thermal equilibrium pressure for each 
density. This curve can be seen in the dashed line of figure \ref{pdfs}.  Note the dashed line only extends 
to densities of $10^2$~cm$^{-3}$  to avoid confusion at higher densities - but the equilibrium curve can be 
seen at higher densities in the distribution itself because the thermal timescales are much smaller than 
any other time scale at those densities, and thus the gas lines up with the equilibrium curve. In the 
``Smooth" simulation, the inflowing gas has a uniform density of $1.0$~cm$^{-3}$ and a thermal equilibrium 
pressure of $4931$~K~cm$^{-3}$. For the ``Clumpy" simulation, the mean inflow density is also $1.0$~cm$^{-3}$, 
yet the flow contains many small clumps of radius $0.55$~pc and a density of $15.2$~cm$^{-3}$, placed  
randomly in a smooth lower background density of $0.25$~cm$^{-3}$. Both the clumps and the low density 
background are in pressure equilibrium at $6857$~K~cm$^{-3}$  and both are stable to thermal instabilities, though they are not in
thermal equilibrium with each other.
The high 
density contrast $\chi=60.8$ between clumps and background results in a filling fraction of $f=0.05$.  
The clump radius was chosen to be much less than the Jeans length at the clump density and pressure 
($L_J=43.3$~pc) so that the clumps would be stable against gravitational collapse.  In both 
runs, the interface between oppositely directed material is initially rippled with a random sequence of 
sines and cosines of amplitude $2$~pc, wavelengths from 40 pc down to $2.5$~pc and power proportional 
to $k^{-3}$.

\section{Results} \label{results}
All of the following analysis focuses on a hockey-puck shaped region that is $40$~pc in diameter and 
$10$~pc thick, centered on the interface between the two flows.  This region is outlined in 
figures~\ref{SmoothPanels} and \ref{ClumpyPanels}.  Figure~\ref{pdfs} shows the joint probability 
distribution in logarithmic density-pressure space for both runs at times $10.1,20.1$, and $27.2$ Myr. 

\subsection{Thermal Properties}\label{thermalprop}
In the Smooth model (left column), material enters the collision region on the equilibrium curve at 
$1.0$~cm$^{-3}$  which coincides with the peak in the equilibrium cooling curve.  As the material 
collides with the oppositely directed material it is initially compressed adiabatically up to the 
flow ram pressure at $1.04\times10^4$~cm$^{-3}$. It then cools and compresses onto the thermal 
equilibrium curve.  With time, more material piles up at higher densities.  Eventually, self-gravity 
takes over at the highest densities and compresses this material further, above the ram pressure 
provided by the flow. At this point, gas collapses and forms a core, or is being accreted by an 
existing core. The core formation or accretion explains the lack of material at densities above 
$\approx 10^{4.6}$~cm$^{-3}$. 

\paragraph{}
The Clumpy flow on the other hand (right column) has material entering the collision region at 
both $0.25$~cm$^{-3}$ and $15.2$~cm$^{-3}$.  Additionally, some mixing occurs between the clumps 
and the background flow which causes the thick band of material below the thermal equilibrium curve.  
At these densities, the thermal time scales are longer than the dynamical time 
[see Fig.~3 of \citet{Heitsch2008b}] - so this material does not equilibrate before colliding with 
the oppositely directed flow.  The low density background appears to also compress adiabatically 
though not to as high pressures as the Smooth run.  At the interface between the two flows there 
are three possible types of interactions due to the two densities present in the flow: 
(a) For background-background collisions, the ram pressure will be $1/4$th of that for the Smooth flow.
(b) Collisions between background material from one side and clumps from the other will produce bow 
shocks equivalent to background material running over a stationary clump with velocity 
$2v_0=16.5$~km~s$^{-1}$, resulting in a ram pressure equal to that in the Smooth model, at 
$1.04\times10^4$~K~cm$^{-3}$.  
(c) Finally, head-on clump-clump collisions, though rare, can produce pressures $15.2$ times the 
ram pressure in the Smooth model. Yet this material will cool fairly quickly due to the high densities.

\subsection{Morphologies}\label{morphologies}
\begin{figure}
\includegraphics[width=\textwidth]{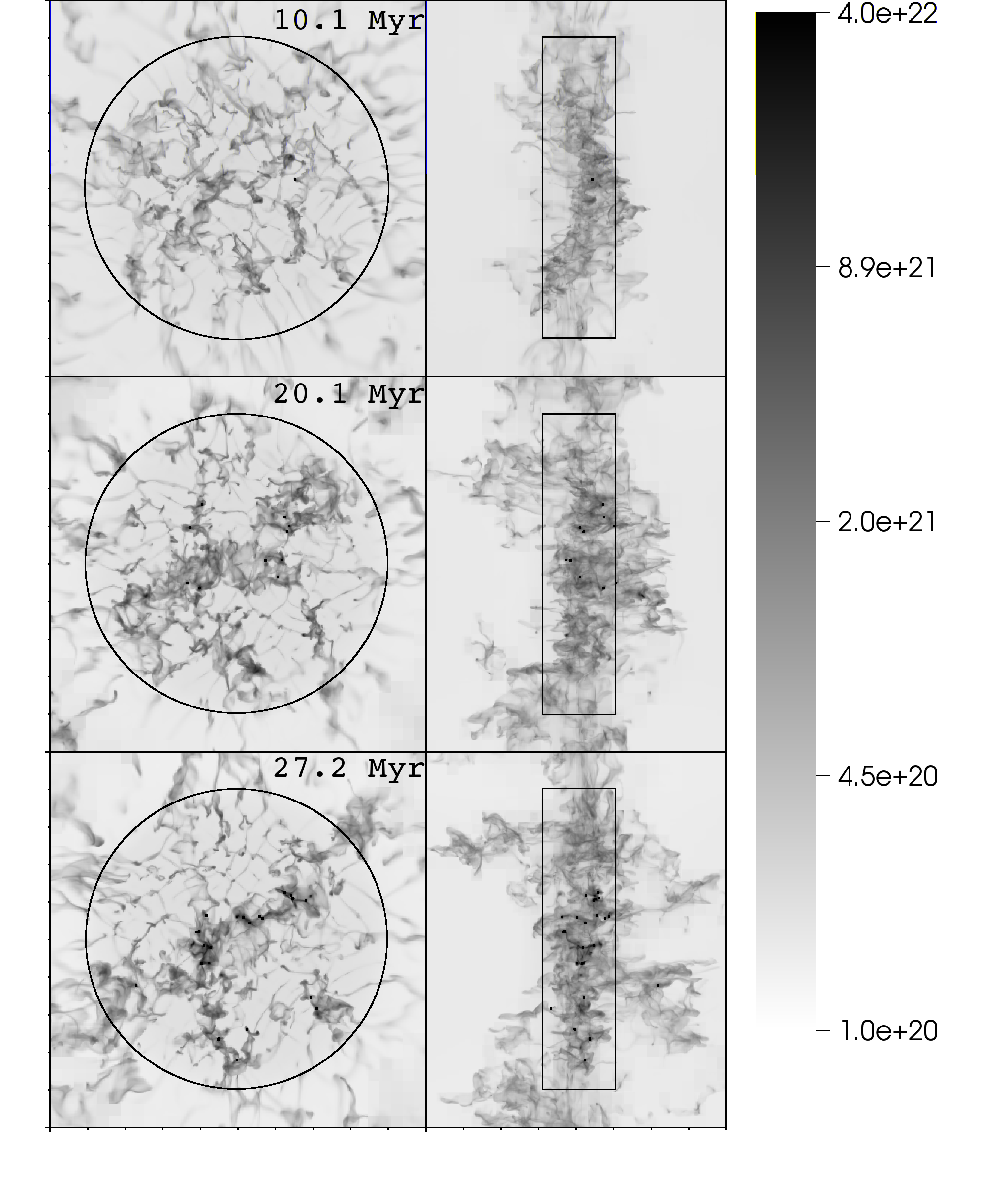}
\caption{Column density in units of cm$^{-2}$ projected parallel (left) and perpendicular 
(right) to flow axis, at $10.1,20.1$, and  $27.2$~Myr for the Smooth run.  Sink particles appear as small black squares. 
Tick-marks are spaced every 5 pc and each panel is $50$~pc tall.}
\label{SmoothPanels}
\end{figure}

\begin{figure}
\includegraphics[width=\textwidth]{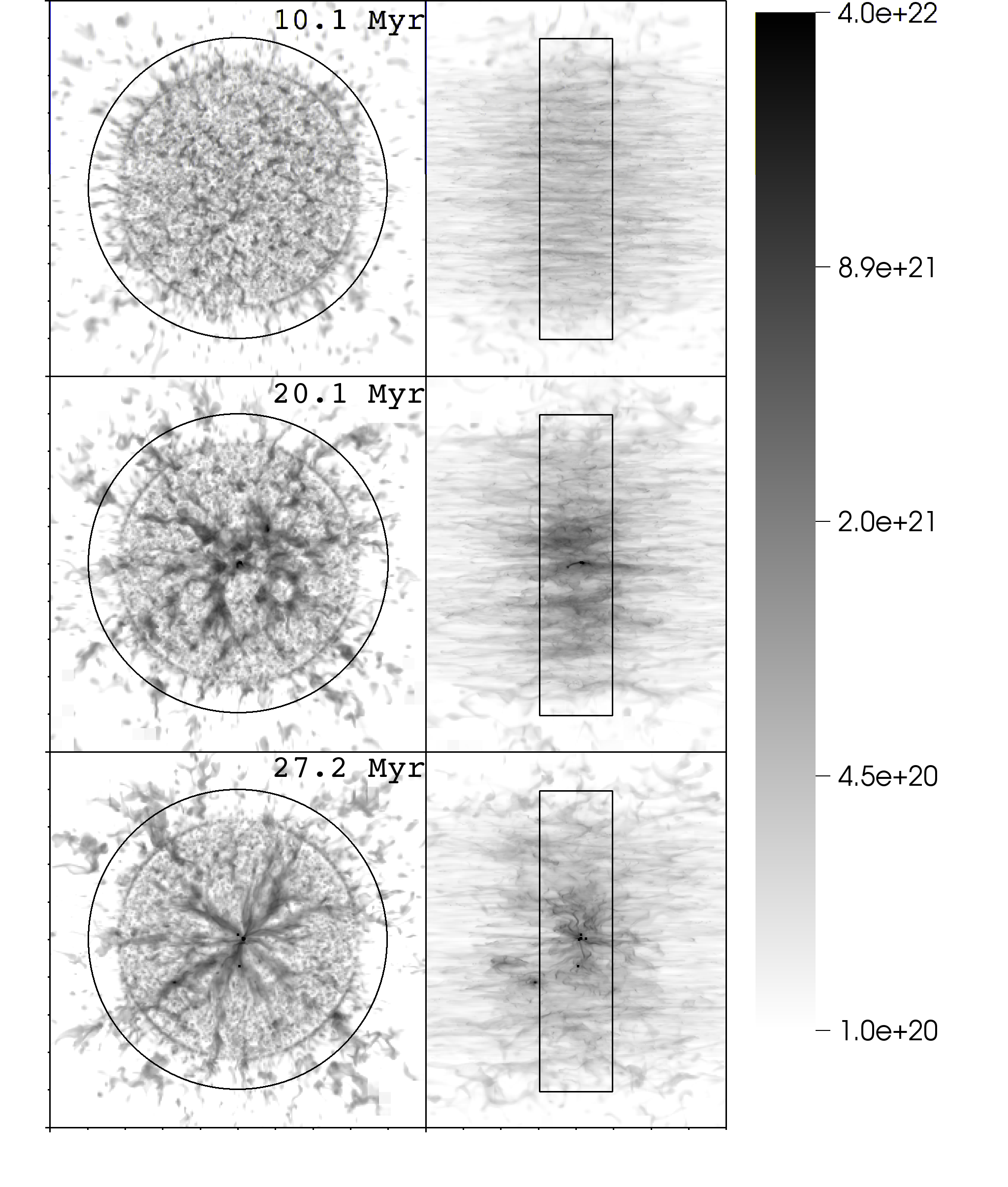}
\caption{Column density in units of cm$^{-2}$, projected parallel (left) and perpendicular 
(right) to flow axis, at $10.1,20.1$, and  $27.2$~Myr for the Clumpy run.  
Sink particles appear as small black squares. }
\label{ClumpyPanels}
\end{figure}

Figures \ref{SmoothPanels} and \ref{ClumpyPanels} show column densities taken along the flow axis 
(left) and normal to the flow axis (right) for the Smooth and Clumpy runs respectively.  Also shown 
are the boundaries of the ``hockey puck'' region used for the following analysis.  The Smooth run 
exhibits the usual filamentary structure due primarily to the non-linear 
thin shell instability [NTSI, \citep{Vishniac1994,Hueckstaedt2003,Heitsch2005,Vazquez-Semadeni2006}], and at
later times gravity. Also visible is material which has been ``splashed'' radially outward from the collision region 
due to the high ram pressures.  The NTSI focuses material into various nodes and by $10.1$~Myr
the first core (solid black square) has formed in one of these nodes \citep[see also] {Heitsch2008a}.  
By $20.1$~Myr, nine cores have formed throughout the complex and by $27.2$~Myr, $27$ cores have begun to 
arrange themselves into clusters.

The morphology of the Clumpy run (Figure \ref{ClumpyPanels}) is quite different from the Smooth run. The view
along the flow axis shows that the clumps are confined to be within the stream by a few pc and they 
have a tendency to cluster around the perimeter.  This is just an artifact of the clump placing algorithm and 
can be safely ignored.  There also tends to be much less radial splashing than in the Smooth run because the 
uniform background component prone to being splashed out is only $1/4$th as dense. The interaction region is
much more extended along the flow axis than in the Smooth run.
Early on, dense clumps run into the lighter 
background and travel a distance before being destroyed.  The timescale for the clump destruction will be of 
order the clump crushing time $t_{cc} = \frac{\chi r_c}{v_w}$ \citep{Klein1994} where $v_w$ is the `wind velocity'
as seen by the clump and $\chi$ is the density contrast.  Since the clump is itself traveling into an oppositely 
directed flow, $v_w=2v_0$ and the distance the clump will travel will be of order 
\begin{equation}
\mathcal{D}= t_{cc} v_0 = \frac{\chi r_c}{2} = 4.3\mbox{ pc}. 
\label{clumpcrushdist}
\end{equation}
If the clump survives for a few
clump crushing times, it will travel distances of $\approx 10$~pc before being destroyed.  This explains 
the more extended interaction region in the upper right panel of figure \ref{ClumpyPanels}.  Later in time, the 
clumps pass through a denser wall of material that has built up and they are also pulled back by gravity - so the 
extent of the collision region shrinks over time.  While the Smooth run has formed nine isolated cores by $20.1$~Myr 
distributed throughout the collision region, the Clumpy run has only just formed a group of three cores near the 
center of the global potential well.  By $27.$~Myr, the entire region is undergoing rapid global collapse and a 
dense group of $20$ cores has formed again near the center of the potential well.

\subsection{Spectra}
To generate spectra we first take a cube of size $d=40.625$~pc, centered on the simulation domain.
The data within the cube is then windowed with
\begin{displaymath}
w(r)=\left\{
\begin{array}{ll}
\cos(\pi r/d) & : r<= d/2 \\
0 & : r > d/2
\end{array}
\right .
\end{displaymath}

The process of mapping AMR data onto a fixed grid often leads to spurious signals at wavelengths 
corresponding to coarser cell sizes.  To reduce this effect we prolongated in Fourier space 
instead of physical space.  The coarse data within the cube was transformed into Fourier space, 
and the various components were then mapped onto a finer Fourier grid.  A phase was then added 
to each Fourier component (since finer cells are not collocated with their coarser parent cells) 
before being transformed back into physical space.  Then this prolongated grid was updated with 
existing data on the next finest level and the process was repeated.  

\begin{figure}
\begin{tabular}{l}
\includegraphics[width=.5\textwidth]{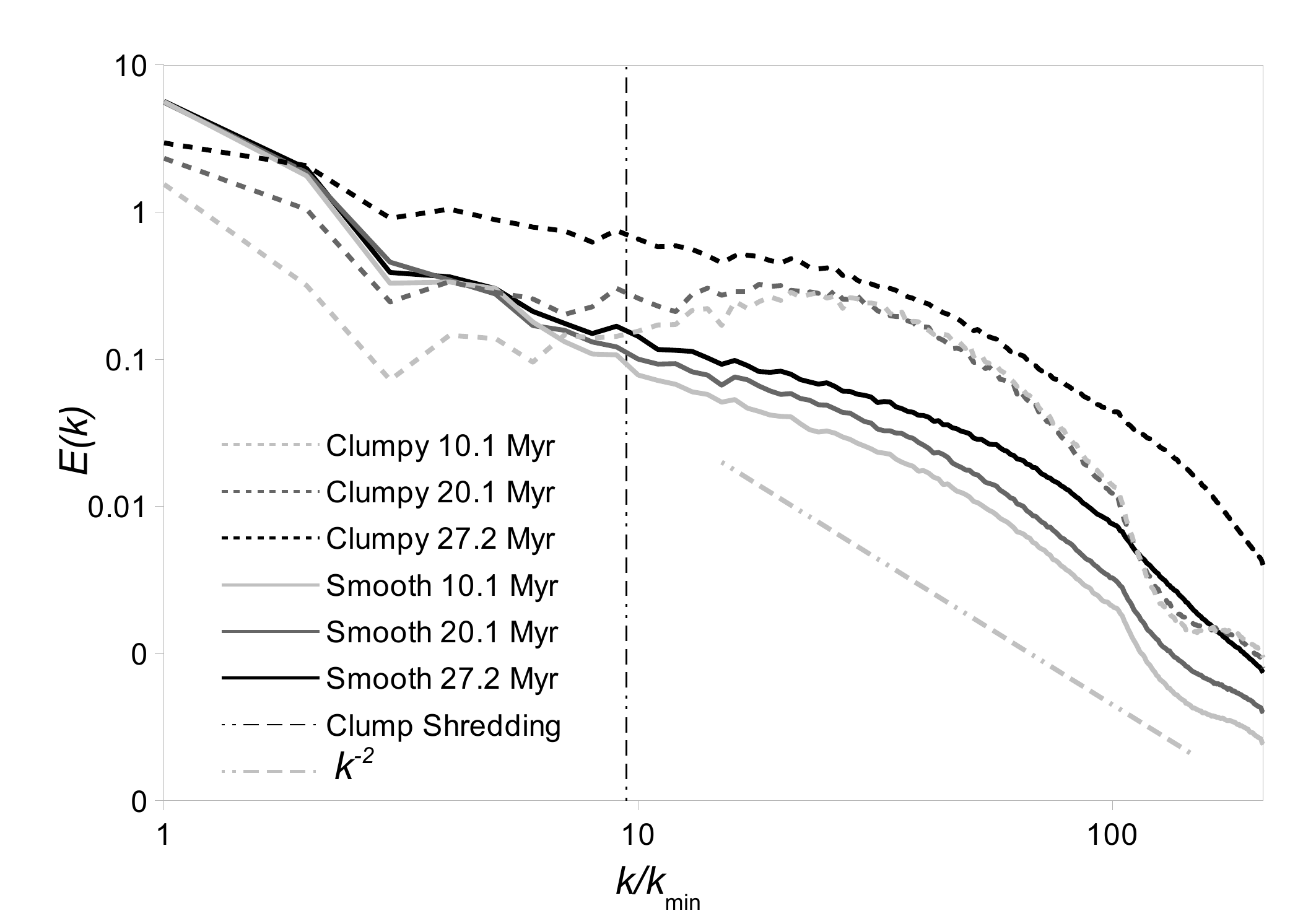}
\includegraphics[width=.5\textwidth]{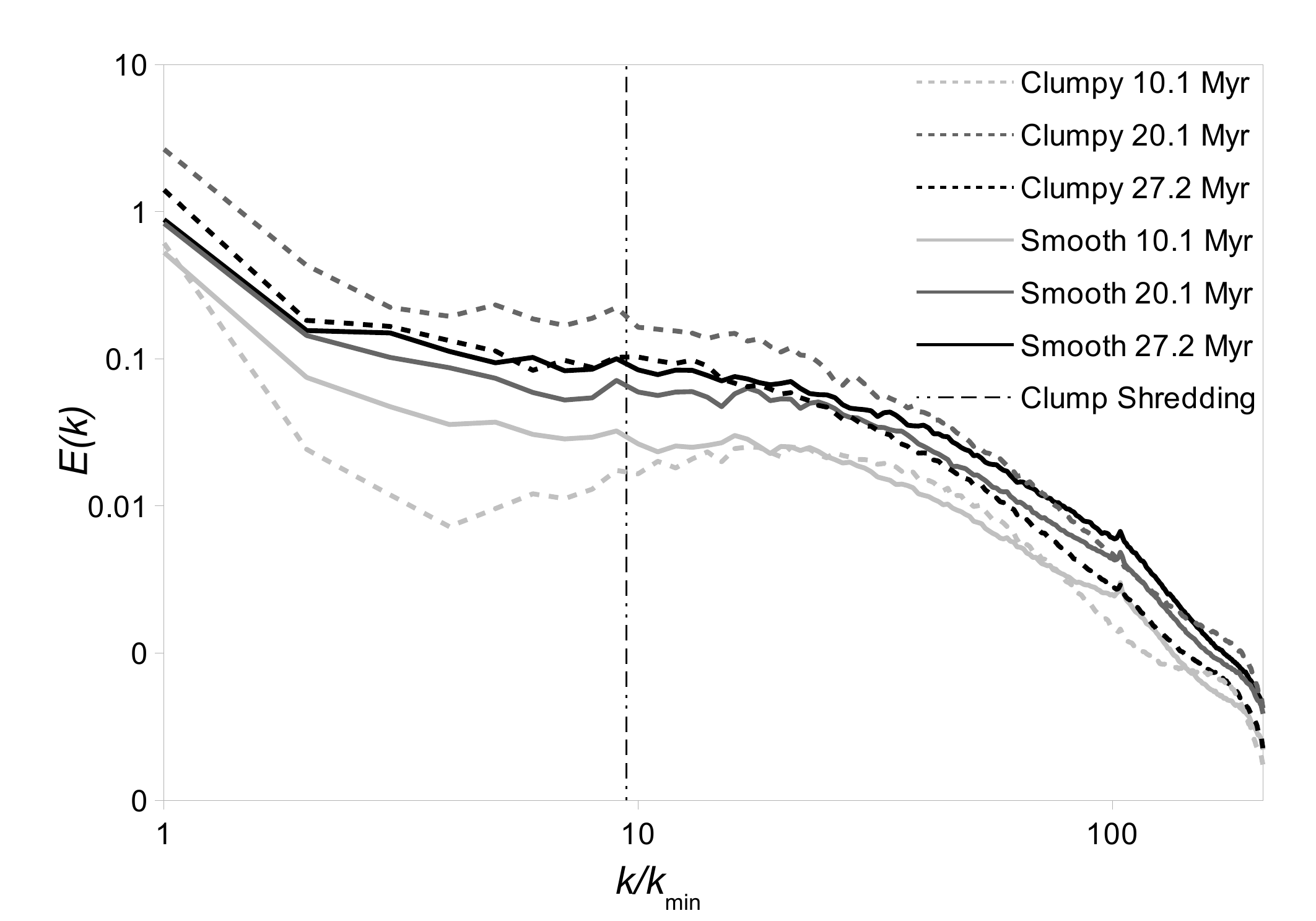} \\
\end{tabular}
\caption{Kinetic energy (left) and gravitational energy spectra (right) for both runs and three times
as indicated in the diagram. The vertical line indicates the clump shredding distance (eq.~\ref{clumpcrushdist}).}
\label{Spectra}
\end{figure}

Figure~\ref{Spectra} shows the kinetic and gravitational energy spectra. At $10.1$~Myr, the 
Clumpy run has an excess of kinetic energy at small scales and a lack of kinetic energy at large 
scales.  While both runs have the same flux of kinetic energy, the Clumpy run contains most of 
the kinetic energy in dense clumps on small scales. When these flows collide, the clumps are able 
to maintain coherence until they have travelled a clump shredding distance (eq.~\ref{clumpcrushdist}) at which point their 
kinetic energy is able to dissipate.  This clump shredding distance can be thought of as a 
`driving scale' for the small scale `turbulence' and roughly corresponds to the break in the power 
spectrum.

By $20.1$~Myr, the Clumpy run has gained large scale energy due to the onset of global collapse, 
and by $27.2$~Myr the Clumpy run has gained energy on all scales due to both the continued global 
collapse as well as the onset of local collapse.  The Smooth run on the other hand does not show 
any growth of large scale kinetic energy, with modest growth of smaller scale kinetic energy.  
This picture is confirmed by the gravitational energy spectra - thought it should be mentioned 
that both spectra are for non-accreted gas.  The overall drop in gravitational energy in the Clumpy 
run between 20.1 and 27.2 Myr is due to the very rapid accretion onto sink particles, following 
the global collapse.

\subsection{Energy budgets}
\begin{figure*}
\includegraphics[width=.49\textwidth]{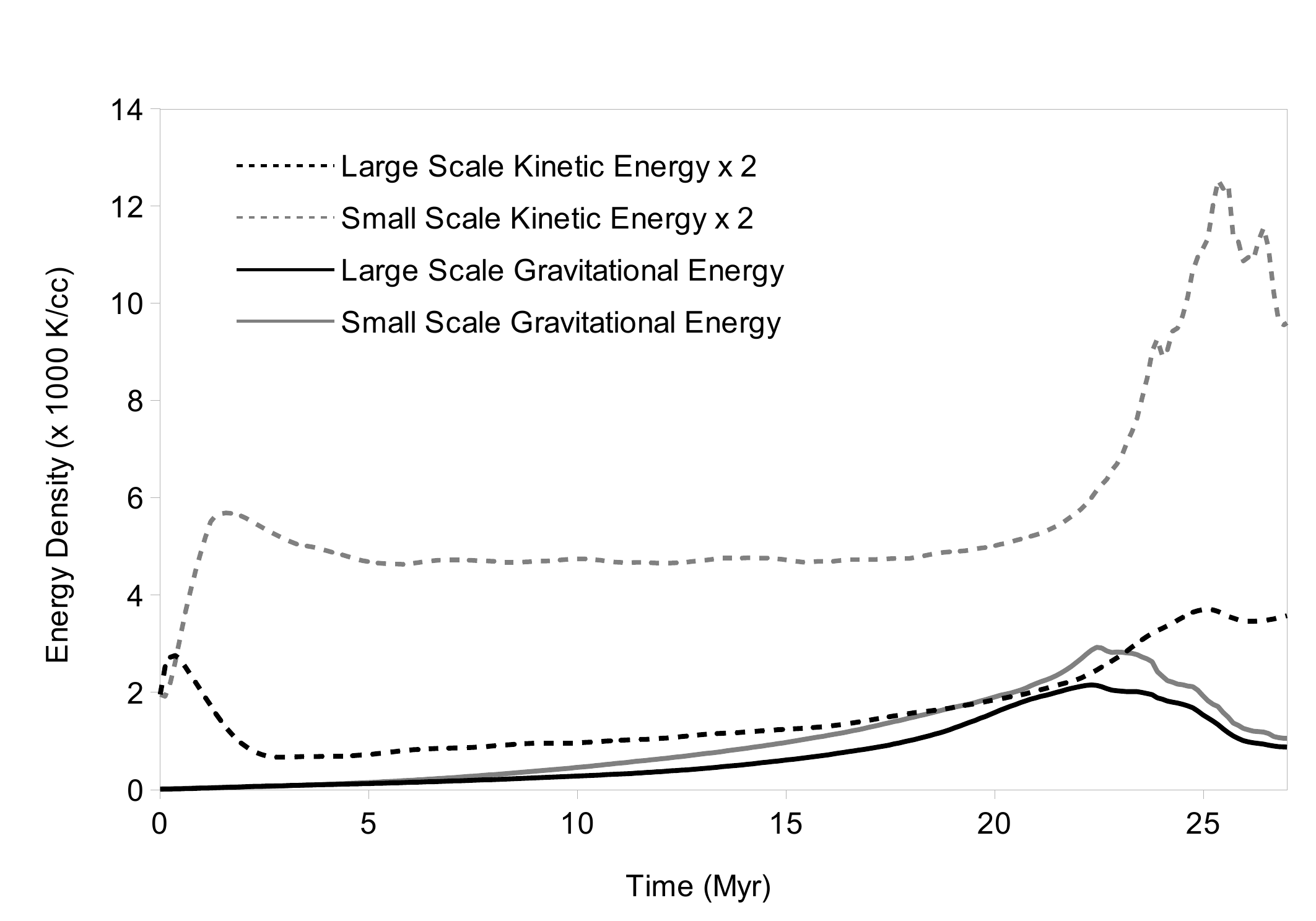}
\hfill
\includegraphics[width=.49\textwidth]{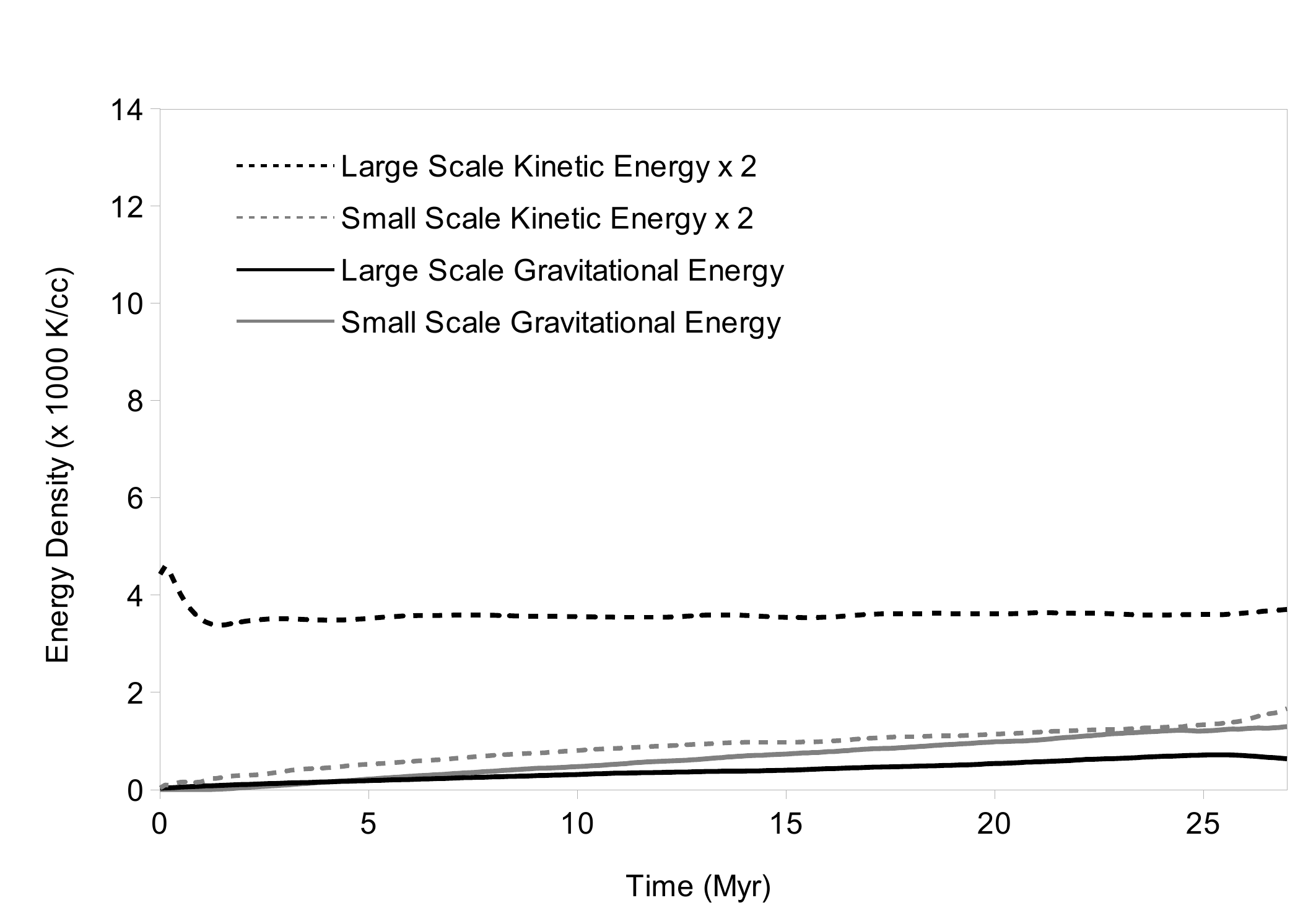}
\caption{Plots showing time evolution of mean kinetic and gravitational energy densities split between 
large ($>\mathcal{D}$, see eq.~\ref{clumpcrushdist}) and small scales for both the Clumpy run (left) 
and the Smooth run (right).}
\label{EnergyBudgets}
\end{figure*}

Figure~\ref{EnergyBudgets} shows the evolution the energy densities at large and small scales.  
We choose the clump shredding distance $\mathcal{D}=4.3$~pc (see eq.~\ref{clumpcrushdist}) as 
the dividing line between ``small" and 
``large" scales.  For the Smooth run (right panel) we note that the kinetic energy on large 
scales is always greater than the large scale gravitational energy. This is consistent with 
the lack of global collapse observed for the Smooth run.  However, the small scale gravitational 
energy becomes comparable to the small scale kinetic energy after $10$~Myr which is when we begin 
to see the formation of isolated cores, indicating local collapse.  The Clumpy run on the other hand shows 
the opposite -- the energy is dominated by small scale kinetic energy which suppresses local 
collapse. Only after $20$~Myr, when the large scale gravitational energy becomes comparable to 
the large scale kinetic energy, do we begin to see global collapse.  By $23$~Myr, much of the 
gas has been accreted into several large cores near the center of the potential well, which 
explains the drop in energies.

While both runs begin with the same total kinetic energy, and while they have the same flux of kinetic 
energy, the Smooth  run is much more efficient at dissipating this energy in large coherent shocks 
resulting in a smaller overall kinetic energy within the collision zone.  In the Clumpy run, the 
density contrast between the clumps and the opposing ambient material leads to a less efficient 
dissipation of kinetic energy.  {\it This excess kinetic energy on small scales suppresses local collapse 
(remember that the clumps themselves are gravitationally stable) but cannot prevent global 
collapse - while in the Smooth run, the higher degree of kinetic energy on large scales resists 
global collapse but not local collapse.} Another way to see this is that the shocks in the Smooth
run will fragment quickly due to the thermal instability. Yet, the velocity dispersion between the
fragments will be small, at least smaller than for the Clumpy run (see Fig.~\ref{MassTotals}, right), 
thus forming structures
that are more or less coherent in velocity space.
Thus, local collapse is seeded. For the Clumpy run, 
thermal instability does not play much of a role, and gas accretion onto the clumps due to cooling
or gravity is negligible within the (dynamical) timescales considered. Thus, local collapse is suppressed,
while global collapse sets in once enough mass in clumps has been assembled.

\subsection{Core Formation}
\begin{figure}
\includegraphics[width=.49\textwidth]{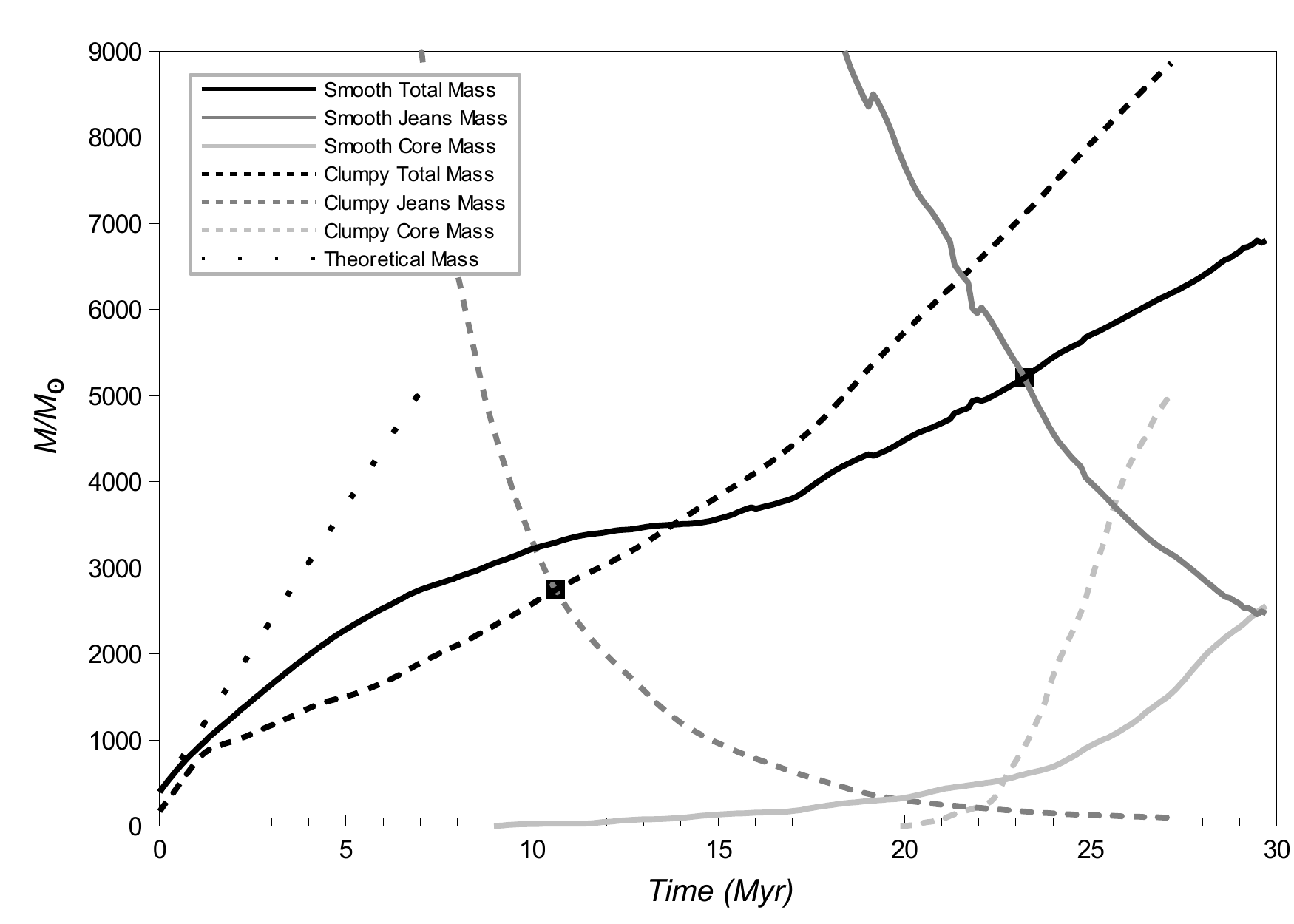}
\hfill
\includegraphics[width=.49\textwidth]{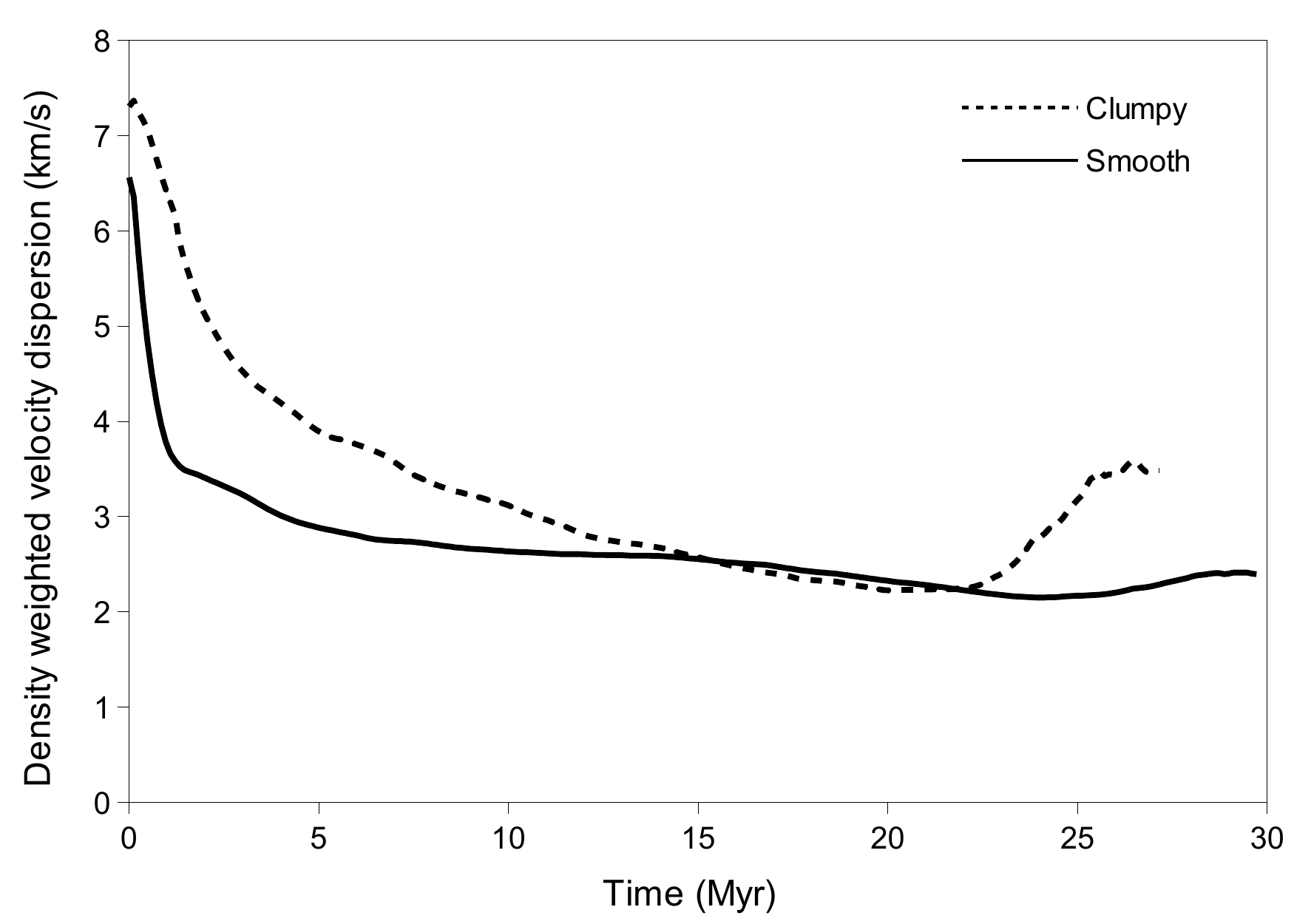}
\caption{{\em Left:} Mass history against time for the Clumpy and Smooth run. Black lines indicate total mass in the 
analysis region, dark gray lines trace the Jeans mass, also in the analysis region, and the light gray lines follow the
mass in sinks, tracing local collapse. The dotted line stands for the mass accumulation expected from simple sweep-up.
Local collapse is suppressed in the Clumpy run until $\approx 20$~Myr, while the Smooth run forms sinks after $\approx 10$~Myr.
{\em Right:} Density-weighted velocity dispersion against time, for the Clumpy and Smooth run, again within the analysis region. 
The Clumpy dispersion is systematically higher until $\approx 15$~Myr, and increases again once global collapse sets in
at $22$~Myr.}
\label{MassTotals}
\end{figure}

The left panel of
Figure~\ref{MassTotals} shows the growth in total mass within the collision region as 
well as the theoretical upper bound (dotted line) using the mass flux $\dot{M}$.  
Both runs collect mass at the inflow rate $\dot{M}$ for the first two Myr, after which
the growth rate drops. For the Smooth run, material is being splashed radially outwards,
and at later times, some of the NTSI fingers develop past the analysis region (see Fig.~\ref{SmoothPanels}). Eventually,
after $15$~Myr, material is falling back in from the edges of the analysis region, increasing
the mass collection rate again.
In the Clumpy run, some of the clumps exit the analysis region on the far side after $2$~Myr.
The overall mass collection rate slowly increases after that, with material falling back in,
and eventually collapsing globally.


The Smooth run begins forming cores at $10$~Myr, and by $25$~Myr, 
the rate of total mass growth and core mass growth have become equal.  
This implies that material is being accreted by the cores at the same rate 
it is entering the collision region.  The Clumpy run (Figure~\ref{ClumpyPanels}) does 
not begin to form cores until $20$~Myr, but then quickly accretes gas at a rate higher 
than the mass flux into the region.  This suggests a degree of global collapse 
not present in the Smooth run.

Since the gas is being compressed and cooled, the Jeans mass at a given sound speed $c_s$ and
a mass column density $\Sigma$
\begin{equation}
  M_J = 1.17\frac{c_s^4}{G^2 \Sigma}
  \label{Jeansmass}
\end{equation}
will drop with time, as shown in Figure~\ref{MassTotals} (left panel).  
It levels out once the minimum temperature of $\approx 10$~K is reached 
(this is only obvious in the Clumpy run, dashed lines, for $t>25$~Myr.
The Jeans mass for the Clumpy run is smaller by at least an order of
magnitude, because of the clumps at higher densities and lower temperatures.
Yet, since these clumps do not form a coherent region with $M>M_J$, local
gravitational collapse is suppressed until $\approx 20$~Myr, and sinks form only once global collapse
sets in, indicated by the increasing slope of the total mass, black dashed line.
The onset of global collapse in the Clumpy run is also visible in Figure~\ref{EnergyBudgets} (left), and
in the velocity dispersions shown in the right panel of Figure~\ref{MassTotals}. 

The Smooth run has a substantially larger Jeans mass that does not level out at a minimum within the model
run time. Yet, because of the rapid local fragmentation into structures larger than a local Jeans mass,
local collapse (and sink formation) sets in at $\approx 10$~Myr. There is no signature of global collapse
in the velocity dispersions, or in the energies.

\begin{figure}
\centering
\includegraphics[width=.5\textwidth]{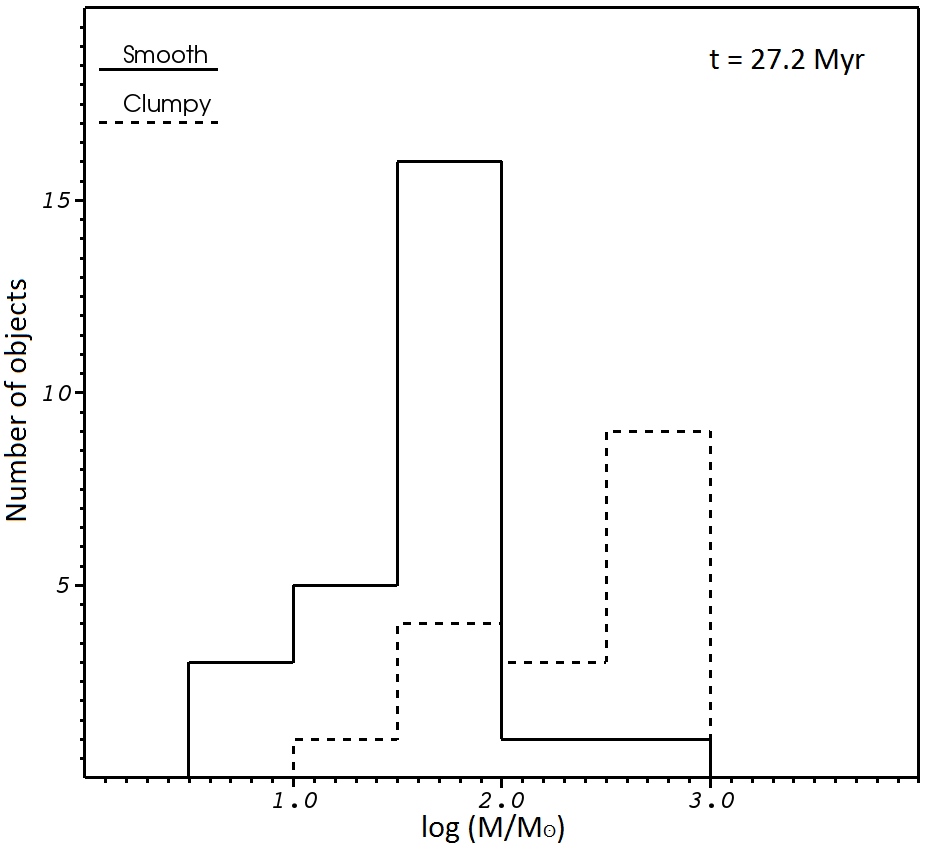}
\caption{Core mass distribution at $27.2$~Myr. Cores in the Smooth run
tend to have less mass than those in the Clumpy run. }
\label{MassHistogram}
\end{figure}

Figure \ref{MassHistogram} shows the distribution of core masses for both runs at $27.2$~Myr.  
The Smooth run forms many intermediate sized cores $<100$~M$_\sun$ consistent with the idea of local 
collapse.  The Clumpy run shows many more high density cores ($100-1000$~M$_\sun$) visible in the center of 
the potential well (Fig.~\ref{ClumpyPanels}) due to global collapse.

\subsection{Mixing}
In both runs, material injected from the left and right side was marked with a tracer ($\rho_L$ and $\rho_R$) 
proportional to the density so that the amount of mixing could be investigated.  We then define a mixing ratio 
\begin{equation}
MR=\frac{2 \min{\left(\rho_L, \rho_R \right)}}{\max{\left ( \rho, \rho_L+\rho_R \right )}}.
\end{equation}
Thus, $MR=0$ indicates the presence of only one tracer (or none - as is the case in the ambient medium 
outside of the flow), and $MR=1$ indicates equal amounts of both tracers with no ambient material mixed in.  
Since we are confining our analysis to the colliding flow region, there should be no ambient material 
present so $\rho=\rho_L+\rho_R$ - and the definition is equivalent to
\begin{equation}
MR=\frac{2 \min{\left(\rho_L, \rho_R \right)}}{\rho_L+\rho_R}.
\label{mr}
\end{equation}

\begin{figure}
\begin{tabular}{l}
\includegraphics[width=.5\textwidth]{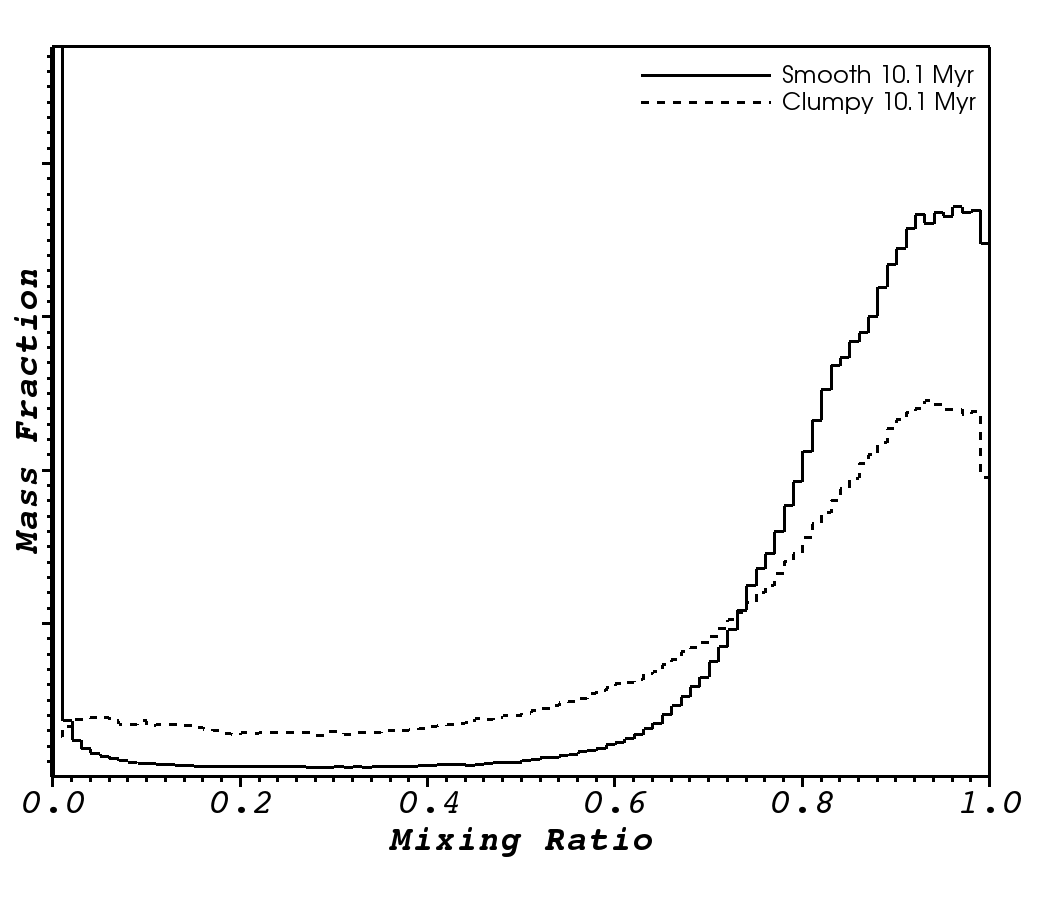}
\includegraphics[width=.5\textwidth]{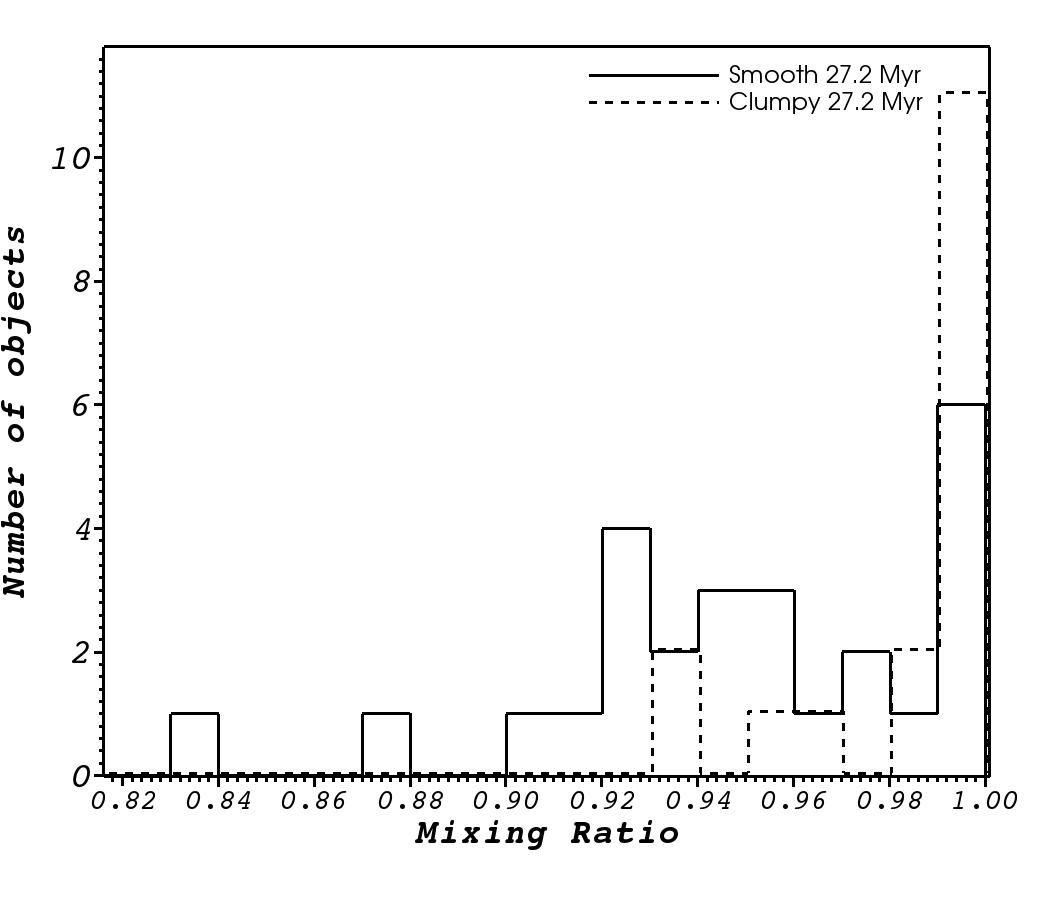} \\
\end{tabular}
\caption{{\em Left:} Mass-weighted mixing ratio (eq.~\ref{mr}) for the gas in the analysis 
region, at $10.1$~Myr. {\em Right:} Mixing ratio of cores at $27.2$~Myr.}
\label{Mixing}
\end{figure}

In the Smooth run, there is a higher mass-fraction of 
well-mixed material (Fig.~\ref{Mixing}, left 
panel; note that the total masses in the analysis region at that time are comparable). 
The Clumpy run tends to have a more spread out distribution of mixing ratios 
than the Smooth run. As clumps drive through the opposing stream - they will shed some of their material 
and provide varying amounts of mixing.  In the Smooth case, the flows interact along a thin interface and 
it is difficult to get unequal amounts of material from either side in the same region.  One might expect 
the Clumpy run to have less well-mixed cores, yet the right panel of figure \ref{Mixing} shows just the 
opposite.  The Smooth run has more cores with lower mixing ratios.  

\begin{figure}
\includegraphics[width=\textwidth]{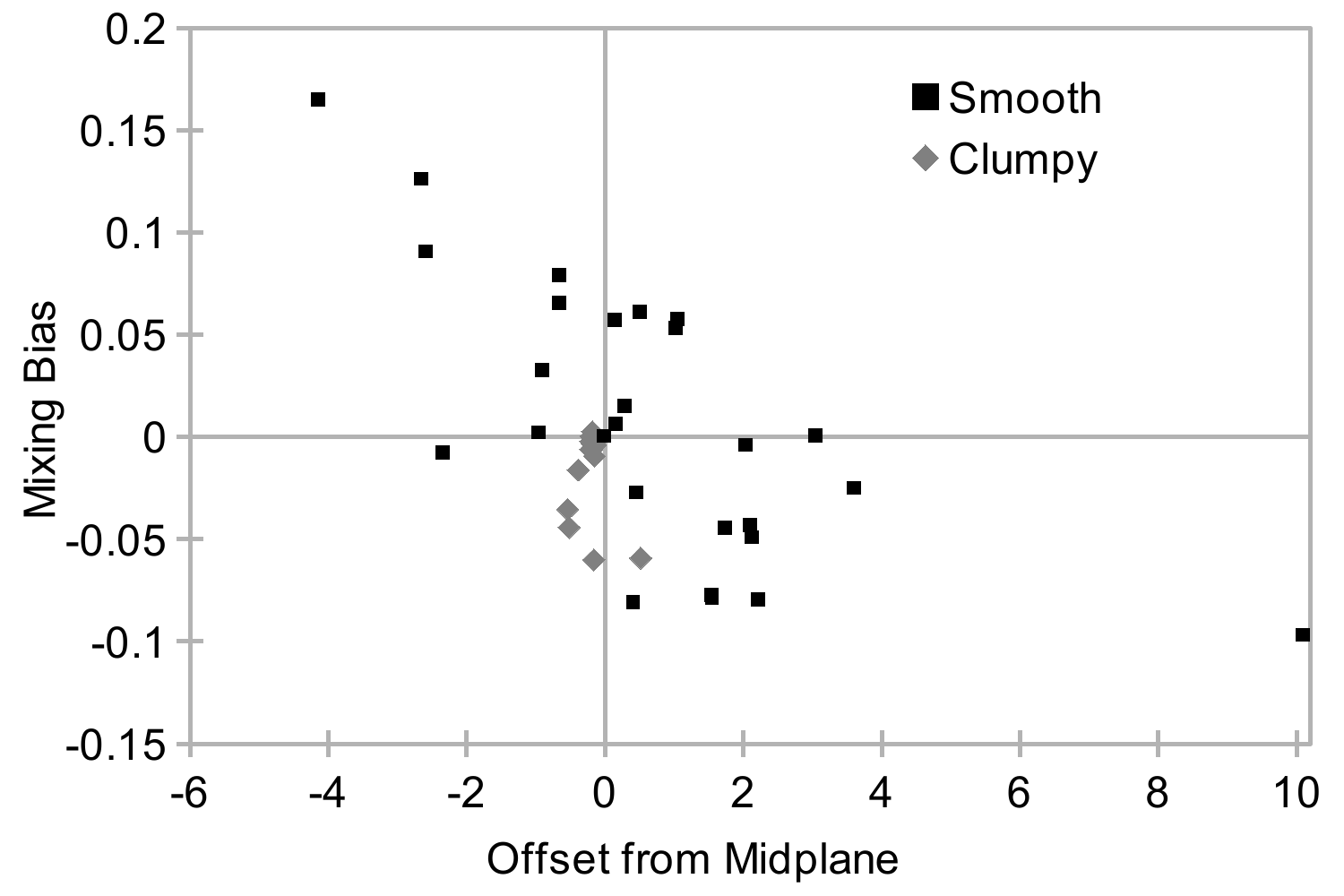}
\caption{Mixing Bias (eq.~\ref{mb}) against distance of cores to mid-plane, at $27.2$~Myr. Smooth cores
show a bias, while Clumpy cores are clustered around $0$.}
\label{MixingBias}
\end{figure}

One possible explanation for this is that the NTSI creates nodes that act to funnel material streaming
into the ``trough'' from only side, while diverting material from the other side.  If so, then 
cores that formed to the left of the collisional mid-plane should be biased towards the right tracer and 
vice versa.  To test this, we define a mixing bias as 
\begin{equation}
MB=\frac{\rho_R-\rho_L}{\rho_R+\rho_L}
\label{mb}
\end{equation} 
and plot it against the core's distance from the y-z mid-plane (Fig.~\ref{MixingBias}). Indeed, 
cores with negative offsets (left of mid-plane) tend to have a higher value for the mixing bias so they 
have more right tracer and are comprised of material primarily from the right side, and vice versa.  
Note that this bias is absent in the cores formed in the Clumpy run, whose cores are clustered around $0$.

%

\section{Discussion and Conclusion} \label{conclusion}
We present the results of two simulations of flow-driven molecular cloud formation. The models are identical
except for the physical conditions of the inflowing gas. One model (``Smooth") starts out with a completely uniform
flow. The other model (``Clumpy") uses the same mass inflow
rate, but the gas is distributed in dense clumplets with a filling factor of $5$\%. This setup is in parts 
motivated by Pringle's et al.~\citeyear{Pringle2001} claim that in order to form molecular clouds from atomic
gas sufficiently fast, dense pockets of already molecular gas should exist in the inflow. Here, we are focusing
on the resulting dynamics of smooth versus clumpy inflows. 

In both cases, the thermal and dynamical fragmentation of the shocked collision region leads to turbulence, at
a level of $10$ to $20$\% of the inflowing kinetic energy. This is consistent with earlier results. The Clumpy
run shows a somewhat higher velocity dispersion initially, since the clumps are less easily stopped in a 
flow collision (see \S\ref{morphologies}). 

Due to the lower compression factor in the Clumpy model, less gas is being cooled to higher densities than in the
Smooth run. Thus, the kinetic energy of the inflow is less efficiently dissipated. Together with a non-contiguous
distribution of cold, sub-jeans, fragments, this leads to a suppression of local collapse for nearly $20$~Myr after
the initial flow collision. At that point, sufficient mass has assembled to induce global collapse of the whole 
region, resulting in a ``star burst" (more appropriately, ``sink burst") at a high sink formation rate.
In contrast, the Smooth run shows local collapse already after $10$~Myr, at less than half the sink formation
rate of the Clumpy run. Due to the local nature of the thermal fragmentation, more fragmentation will occur
with increasing resolution \citep{Hennebelle2007}, thus, the times quoted here are {\em upper} limits for the onset
of local collapse. Nevertheless, structured flows can delay the onset of (substantial) local collapse.
Global collapse is only clearly visible in the Clumpy run.

The differences between Clumpy and Smooth inflows extend to the mixing efficiencies. Somewhat counter to 
a naive expectation, the Smooth initial conditions result in less well mixed material (and cores). This is
primarily due to the NTSI funneling material preferentially into the troughs located far into the opposing
inflow. For the Clumpy run, the global collapse of the accumulated clumps erases all memory of the initial
inflow direction.

Obviously, we have chosen two extremes as our initial conditions. It is more likely that the inflows themselves
will contain turbulent velocity and density structures that are coherent in space. Spatial coherence leads to
stronger shocks in the collision region, and thus to more efficient energy dissipation. In that sense, our Clumpy
run is overestimating the effect of structured inflows.  

\acknowledgments{}
Support for this work was in part provided by NASA through awards issued by JPL/Caltech through Spitzer 
program 20269, the Department of Energy through grant number DE-SC-0001063, Cornell University through 
agreement number 41843-7012, the National Science Foundation through grants AST-0807363 as well as the 
Space Telescope Science Institute through grants HST-AR-11250 and HST-AR-11251.  We also thank the 
University of Rochester Laboratory for Laser Energetics and funds received through the DOE Cooperative 
Agreement No. DE-FC03-02NA00057. FH acknowledges support from NSF grant AST-0807305, NHSC 1008 and the 
NC Space Grant young investigator program.  
This research was also supported in part by the Center for Research Computing at the University of 
Rochester as well as the National Science Foundation through TeraGrid resources provided by the 
National Center for Supercomputing Applications.

\end{document}